\documentclass[times,review,preprint,authoryear]{elsarticle}

\usepackage{jasr}
\usepackage{framed,multirow}
\usepackage{array}
\usepackage{amssymb}
\usepackage{latexsym}

\usepackage[switch]{lineno}
\usepackage{url}
\usepackage{xcolor}

\definecolor{newcolor}{rgb}{.8,.349,.1}
\definecolor{lg}{rgb}{0.95,0.75,0.0} 

\usepackage[citebordercolor=white]{hyperref}

\journal{Advances in Space Research}

\begin{document}

\verso{Plutino \textit{et al.}}

\begin{frontmatter}

\title{A new catalogue of solar flare events from soft x-ray GOES signal in the period 1986-2020}%

\author{Nicola Plutino}
\author{Francesco Berrilli\corref{cor1}
}
\cortext[cor1]{Corresponding author:}
\ead{francesco.berrilli@roma2.infn.it}
\author{Dario Del Moro}
\author{Luca Giovannelli}

\address{Department of Physics, University of Rome Tor Vergata, Via della Ricerca Scientifica, 1, Rome 00133, Italy}

\received{30 March 2022}
\finalform{}
\accepted{}
\availableonline{}

\communicated{}
\begin{abstract}
Solar flares, along with other sun-originated events such as Coronal Mass Ejections, fast solar wind streams, and solar energetic particles are among the most relevant events in Space Weather.
Moreover, solar flares are the most energetic processes that occur in our solar system.
The in-depth study of their occurrence statistics, both over extended periods or during individual solar cycles, allows us to improve and constrain the basic physical models of their origin.
Increasing the number of detected events, especially those of lower intensity, and the number of physical parameters that describe the detected flares is, therefore, a mandatory goal.

In this paper, we present a computationally efficient algorithm for the detection of solar flares in the soft-X solar flux provided by the GOES (NASA/NOAA) satellite constellation. 
Our code produces a new flare catalogue increasing the number of events with respect to the official GOES list. 
In addition to increasing the number of identified events, the catalogue contains information such as: an estimate of the total energy released, start and end time of the event, possible overlap with other events, background level of the GOES X-ray emission close to the revealed event.
After a detailed description of the detection algorithm, we carry out a preliminary analysis of the flares reported in our catalogue and compare our results with the official list of GOES for the period from 1998 to 2020.
\end{abstract}


\end{frontmatter}



\section{Introduction}
\label{section:introduction}
Space weather events (SWEs), in the form of solar flares, Coronal Mass Ejections (CMEs), fast solar wind streams, solar energetic particles (SEPs), accelerated by magnetic reconnection processes during flares and CME-driven shocks,
occur regularly albeit with different frequencies. All these SWEs affect the physical state of the Earth's upper atmosphere, i.e., the magnetosphere, the ionosphere and the thermosphere, and therefore they are important both in defining the physics of Sun-Earth relations and because they have significant operational benefits for satellites in low Earth orbits (LEOs).

The many open questions regarding the physics governing solar activity, the difficulty of predicting instability in solar plasma or identifying the magnetic feature connected to SWEs, including flares \citep[e.g.][]{park2020comparison,cicogna2021flare}, have led the space agencies to focus their efforts on Space Weather phenomena and precursors \citep{esa_ssa2010,Crown2012,Murray2017,Ishii2017,Berrilli2019,Plainaki2020}.

Solar flares release a huge amount of energy,
usually $10^{20}$~J, however the largest solar flares can reach energies of around $10^{25}$~J, sometimes up to $10^{26}$~J, over a timescale of hours \citep{schwenn2006space}.
These impulsive phenomena may trigger spectacular SWEs effects, starting from the historical events that opened the study of the Sun-Earth relationship, such as the Carrington Event of September 1859 \citep[e.g.][]{Smart2006} or the great aurora of February 1872 \citep[e.g.][]{Berrilli2022}.

Solar flares and CMEs do sometimes occur at the same time – and the strongest flares are almost always correlated with CMEs – but the exact relationship between the two events has been long debated.
The late idea that CMEs were initiated by large flares \citep{lin1976non}, has been  disputed by \cite{zhang2001temporal} and \cite{zhang2006statistical} which showed that CMEs are statistically initiated before the onset of the associated flares. 
Recent works, making use of space-borne observatories and computational models,  effectively investigated the solar surface and atmosphere associated with a flare and CME to analyze the dynamics of the magnetic field before and during the ejection event. Among others, a recent work by \cite{temmer2017flare} confirmed that magnetic reconnection processes associated with the CME “were already ongoing before the start of the impulsive flare phase”.

Since it is not the purpose of this work to go into the details of the single flare model or of the magnetic instability processes, responsible for the flare triggering of active regions, we refer to specific literature to deepen these aspects \citep[e.g.][]{Priest1981,Tork2004,Fletcher2005,Pucci2014, Toriumi2019}.

Rather, in this work we focus on the need to increase the statistics of flares in order to investigate the nature of physical processes, and the consequent models, necessary to explain their distribution over time and the physical nature that connects them to the solar magnetic cycle \citep[e.g.][]{Vlahos1995,Boffetta1999,Priest2002,Charbonneau2001,Viticchie2006,aschwanden_algo,Sadykov2019,Lu2021}.
In literature, several papers presented the analysis of the distribution of soft X-ray peak fluxes of solar flares suggesting that power laws with slope with a coefficient $\alpha = 1.6-2.2$ are a good estimate of this behavior \citep[e.g.][]{crosby1993,Boffetta1999,Veronig2002,aschwanden_algo, li2016}, although other functions, i.e., a lognormal distribution, have recently been proposed \citep[e.g.][]{Verbeeck2019}.

Indeed, as shown in \cite{aschwanden_algo}, there are still many open and intriguing questions concerning the statistics of solar flares.

The statistical analysis of flares during the different solar cycles, and possibly within the same cycle, requires to increase the detection of events as much as possible and to determine different characteristic parameters of X-ray emissions. These physical parameters are for example: an estimate of the total energy emitted, since the peak flux alone provides a partial information of the  energy budget of the flare in the observed spectral range, the multiplicity, which is linked to the release efficiency of the active region where the flare is located, or the start and end times of the flare from which to extract the waiting times, the distribution of which makes it possible to discriminate between different models, i.e., shell model for MHD turbulence \citep[e.g.][]{Giuliani2000} or fractal-diffusive self-organized criticality \citep[e.g.][]{aschwanden_algo}.

With this purpose in mind, we have developed an algorithm which, based on GOES SXR light curves and on the evaluation of the background level of the GOES X-ray emission close to the revealed event, estimates several of these parameters that are used to populate the catalogue.
This result allows us to carry out a preliminary statistical analysis on our data. In particular, an analysis was performed on the 1998-2020 events in order to study the distributions of peak fluxes under different analysis conditions. For a direct comparison of the distributions of the peak fluxes between us and GOES, and especially of the distributions when we consider the evaluation of the background level of the GOES X-ray emission close to the identified event.

To achieve this goal we use as input data of our algorithm the long channel (1-8 \AA) in the soft X-ray (SXR) range because of the higher signal-to-noise ratio and better sensitivity to low-intensity events (available at \url{https://www.ngdc.noaa.gov/stp/satellite/goes/dataaccess.html}).

To implement our algorithm we have relied on the algorithm proposed in \cite{aschwanden_algo} by modifying some procedures and parameters (for details about the comparison between our algorithm and that of \cite{aschwanden_algo} see appendix A and the discussion in the next sections) but above all by adding new physical parameters.
Indeed, compared to \cite{aschwanden_algo} work, we have increased the number of physical parameters (attributes) of the revealed events. In fact, as we have discussed earlier, the selection of the most appropriate model to describe the flare distribution requires different parameters from the peak flux alone.

Using GOES SXR light curves in the period 1986-2020 we reveal a total of 334123 events that will form the population of our catalogue. The whole catalogue containing the event attributes estimated by our algorithm for all the identified flares are made available through the web page \url{https://github.com/nplutino/FlareList} together with Python scripts to retrieve selected data that match user-defined criteria.

\section{GOES data set}
GOES consists of a series of geostationary satellites (orbiting the Earth at a height of $35,790$ km), which produces an uninterrupted time series of solar X-ray fluxes, together with meteorological observations of the Earth. This continuous stream of data is guaranteed by an overlap in time so that there are always one to three spacecraft recording data. 

GOES X-ray Sensor (XRS) data have been recorded by NOAA satellites since 1974 and are archived at the NOAA National Center for Environmental Information (NCEI). Each GOES satellite is equipped with two X-ray sensors which provide solar fluxes for the wavelength bands of 0.5-4 \AA ~(short channel) and 1-8 \AA ~(long channel).

NOAA Space Weather Prediction Center (SWPC) has used this data to produce the 1-minute and 5-minute averaged X-ray data sets. Moreover, it is responsible for producing an official X-, M-, and C-class solar flare catalog, observed by the 1–8\AA ~channel of the GOES XRS, which we will be referred to as the GOES catalog from now on. This catalog, derived from an automatic procedure applied to XRS lightcurves, is the gold standard for studies on the statistical properties of flares. Given its relevance, we will describe in an extremely synthetic way, the standard NOAA algorithm applied to the XRS signal and which define a GOES flare event.

SWPC algorithm populates GOES catalogs with the following flare detection criteria:

\begin{enumerate}
    \item a \textit{GOES X-ray flare event} starts when 4 consecutive 1-minute X-ray flux are above the B1 threshold, i.e. above a flux threshold of $10^{-7}$ W/m\textsuperscript{2};
    \item all 4 flux values monotonically increase;
    \item the fourth flux value is greater than $1.4$ times the first value in the sequence of 4 data points.
\end{enumerate}

\noindent Solar flares are classified (A-B-C-M-X scale) taking in consideration the flux in the soft-X band at the peak of the flare emission.
The event ends when the flux, during the decaying phase, drops below half of the peak flux. 
The data set used in this work covers the period from 1986 to 2020 (three solar cycles), with data from GOES-6 to GOES-16.
Data from GOES-6 to GOES-15 have been downloaded in their February 2021 version (\url{https://www.ngdc.noaa.gov/stp/satellite/goes/dataaccess.html}), while GOES-16 data have been downloaded in their March 2021 version (\url{https://www.ngdc.noaa.gov/stp/satellite/goes-r.html}). However, as described in the GOES XRS Operational Data document,
(\url{https://www.ngdc.noaa.gov/stp/satellite/goes/doc/GOES_XRS_readme.pdf}), to get true fluxes for data up to GOES-15, users must remove the SWPC scaling factors. Dividing the long band flux by 0.7, such corrected fluxes and corresponding flare indices will agree with those obtained by GOES-16. We applied this correction to the GOES light curves, analyzed by our procedure, in order to obtain true fluxes for the entire list of events.
At the same time, since data is recorded from different satellites, data points in the whole series do not show the same temporal resolutions: until 2010 data have a temporal resolution of 3s, while more recent data have a 2s resolution. In addition, data present gaps and instrumental spikes which must be handled before the event detection itself.
For this reason, data manipulation starts with a smoothing process.

\section{Flare detection algorithm}
In this section we describe our mining procedure applied to GOES soft X-ray light curves to identify X-ray events (i.e., flares). The methodology includes a data pre-processing step, preparatory to the identification of events, and subsequent steps in which the events are identified and different physical quantities associated with the events are calculated.
For the sake of clarity we divide this description in subsections, starting with the data pre-processing, and the localization of x-ray emission local minima and maxima, which is critical for our definition of flare event. The last two subsection are devoted to the handling of overlapping flare events and to the description of the outputs of the flare search algorithm. In subsection 3.3 we discuss how to deal with the pre-flare background. The estimation of this flux, which includes not only the contribution of the active region hosting the event, but the integrated flux over the entire solar disk (sun-as-a-star) and beyond the limb, is necessary to maximize the efficiency in detecting flare of a minor class and in correctly estimating the energy associated with the identified event.
\label{section:dataprocessing}
\subsection{Data pre-processing: smoothing procedure}
An initial smoothing procedure is applied to the XRS signal as a one-dimensional data noise reduction. 
Data smoothing techniques for GOES soft x-ray light curves have been described extensively \citep[e.g.][]{aschwanden_algo}.
We replicate some of these procedures using Python libraries and analysing 24 hours of data at a time. The use of 24 hour batches was decided to facilitate the execution of the procedure, but introduces some possible problems if a flare occurs near the end of the batch and extends after the end of the day. At the moment we neglect this fact for two substantial reasons: first statistically no more than 4\% of the events occur in the last hour of the day, second, since the duration of the event depends on its intensity \citep[e.g.][]{Kashapova2021} only the flare more energetic that occur around 24:00 UTC extend significantly after the end of the day leading to a significant underestimation of descriptors such as total energy. We checked all the identified class X events and we report that of the 528 flares recorded for the entire 1986-2020 period only 17 have a peak after 23:00 UTC (i.e., 3.2\%) and only two events have a peak near the midnight. A procedure that considers these few cases, taking into account the initial hours of the following day, has been implemented and will allow to correct the values of total energy and end time of the flare. In any case, since in the present work we will carry out the analysis of peak fluxes only, this problem has no effect.

To deal with the varying temporal resolution of the data we perform a temporal sub-sampling by replacing a set of data points with their average flux.
Since the typical solar flare duration is in the range of minutes \citep[e.g.][]{reep2019determines}, this operation can be performed without significant loss of information.
This procedure not only produces a uniform time-spaced data set, but also reduces the number of data points to be processed by the algorithm, improving efficiency. 
The sub-sampling operation is performed using the function \textit{resample} of the module \textit{pandas} for Python \citep{pandas}. 
This function correctly re-samples our data every 12s and is adept at excluding data gaps under typical conditions.

The main purpose of the smoothing procedure is to mitigate or eliminate the effect of instrumental spikes on the light curve to be analyzed. Indeed, sudden flux increases of instrumental origin, i.e., non-solar origin, can produce a false flare detection.
This problem has already been pointed out and solved by introducing a parameter that describes how fast the slope of the light curve changes \citep{aschwanden_algo}. 
We use a moving time window, with a time window size of 60s, which associates each data point with a value 
\begin{equation}
Q = F_{max}/F_{min}   
\end{equation}
\noindent where $F_{max}$ and $F_{min}$ are the maximum and minimum flux value in the time window.  Therefore, the Q-value for each data point contains information on the neighbour of that data point and on how fast the flux changes in this time interval.
Subsequently, the algorithm performs a classification procedure. Any point with a value of $Q > Q_{crit}=10$ is identified as a spike and removed from the data set.
We tested this method with data from the year 1987, one of the years greatly affected by the spike problem. The spike detection resulted in a total of 4968 events, less than the number of data points in a single day.
Therefore, it should be emphasized that if on the one hand the elimination of these instrumental spikes must be carried out to eliminate false identifications, on the other hand the effect on the total statistics is in practice negligible.

\begin{figure}
\centering
\includegraphics[width=0.4\textwidth]{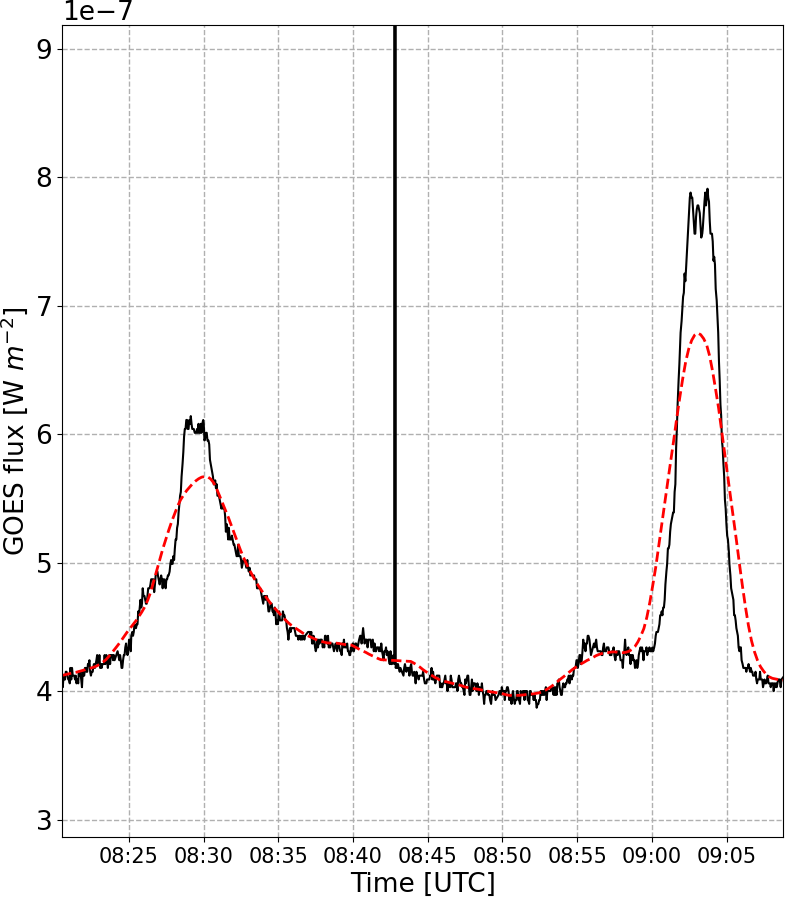}
\caption{Data smoothing. The black continuous line represents the raw data. A major data spike can be observed at about 8:42 UTC. The red dashed line represents the signal obtained after the smoothing operations. Data refer to 2\textsuperscript{nd} November 1987.}
\label{fig:smooth}
\end{figure}

The last step of the smoothing procedure is a boxcar average operation, which is crucial to improve algorithm efficiency. Indeed, our algorithm computation time strongly depends on the number of maximum and minimum points to be analysed. With a smoother curve, we are automatically excluding maximum and minimum points coming from signal noise. 
Given the boxcar size (the number of data points $2n+1$ to be used in the average procedure), fluxes from $x_1$ to $x_{2n+1}$ are summed up and the total flux is then divided by $2n+1$. The boxcar size is 21 data points, corresponding to approximately 4 minutes. The flux obtained this way is attributed to the middle point $x_n$. Then the average operation moves on points from $x_2$ to $x_{2n+2}$. We can describe the whole boxcar averaging operation with the recursive formula:
\begin{equation}
    x_i = \frac{1}{2n+1}\sum_{j=i-n}^{i+n} x_j
\end{equation}
All the steps described so far produce the smoothed data sets. In Figure \ref{fig:smooth} an example of the smoothing procedure on real data is shown.

Once this pre-processing of the data has been completed, it is possible to move on to identifying the events. The flare detection algorithm will be described in the next sections.
\subsection{Flare identification: detection of local maximum and minimum}
\begin{figure}
\centering
\includegraphics[width=0.4\textwidth]{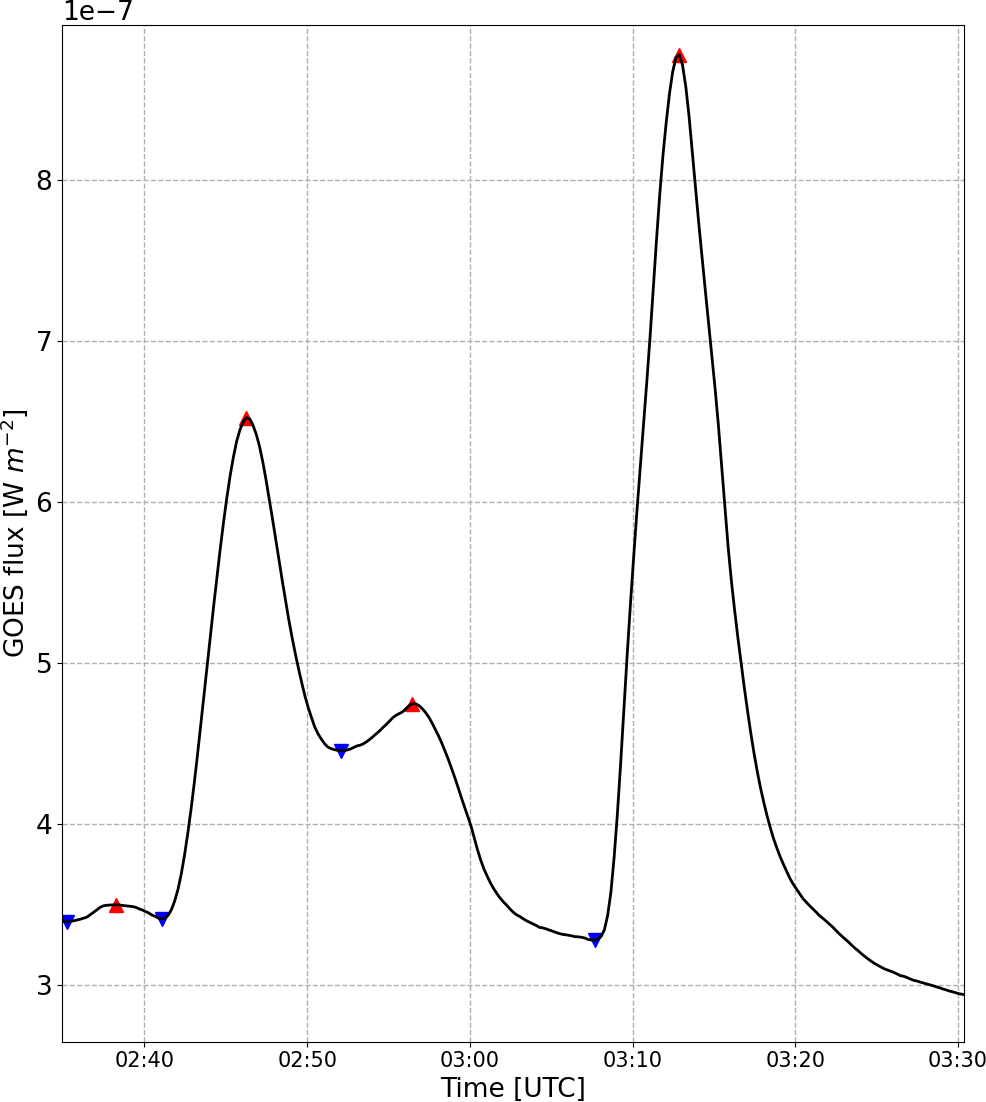}
\caption{Maximum and minimum detection procedure. Red upward triangles mark maximum values, while blue downward triangles mark minimum values. Data refer to  2\textsuperscript{nd} November 1987 GOES XRS signal.}
\label{fig:maxmin}
\end{figure}

The identification of a flare event starts with the detection of local maxima and minima. More in detail, local minima will be taken as possible event start points, while local maxima will be used as candidates for flare peak flux values.
Indeed, GOES light curves present small variations even after the smoothing procedure has been applied. For this reason, the search for the maximum (or minimum) is carried out in a moving sub-sample of 5 points.
A data point is considered maximum (minimum) only if its value is higher (lower) than the values of the two points on its left and of the two points on its right.
This operation considerably reduces the number of extremes that the algorithm has to analyze to identify events. An example of this procedure is shown in Figure \ref{fig:maxmin}.
\subsection{Flare identification: definition of an event}
Once all minimum and maximum values are listed, a new step of the flare identification procedure searches for candidate flare events.
However, as reported in the previous sections, the soft X-ray flux level background must be taken into account in order to identify event candidates. As already reported in the paper by \cite{Wagner1988}, the flux integrated soft X-ray produced by all the active regions on the whole solar disk (sun-as-a-star) and beyond the limb changes continuously and is also modulated by the solar activity cycle. This flux level can be greater than the expected flux for flare of a minor classes (A or B), and during the maximum of activity also of C-class flares. Particularly important in this context is the work of \cite{Ryan2012} which, with the aim of deriving flare thermal properties, introduces an automated temperature and emission measure-based background subtraction method.

\begin{figure}
\centering
\includegraphics[width=0.4\textwidth]{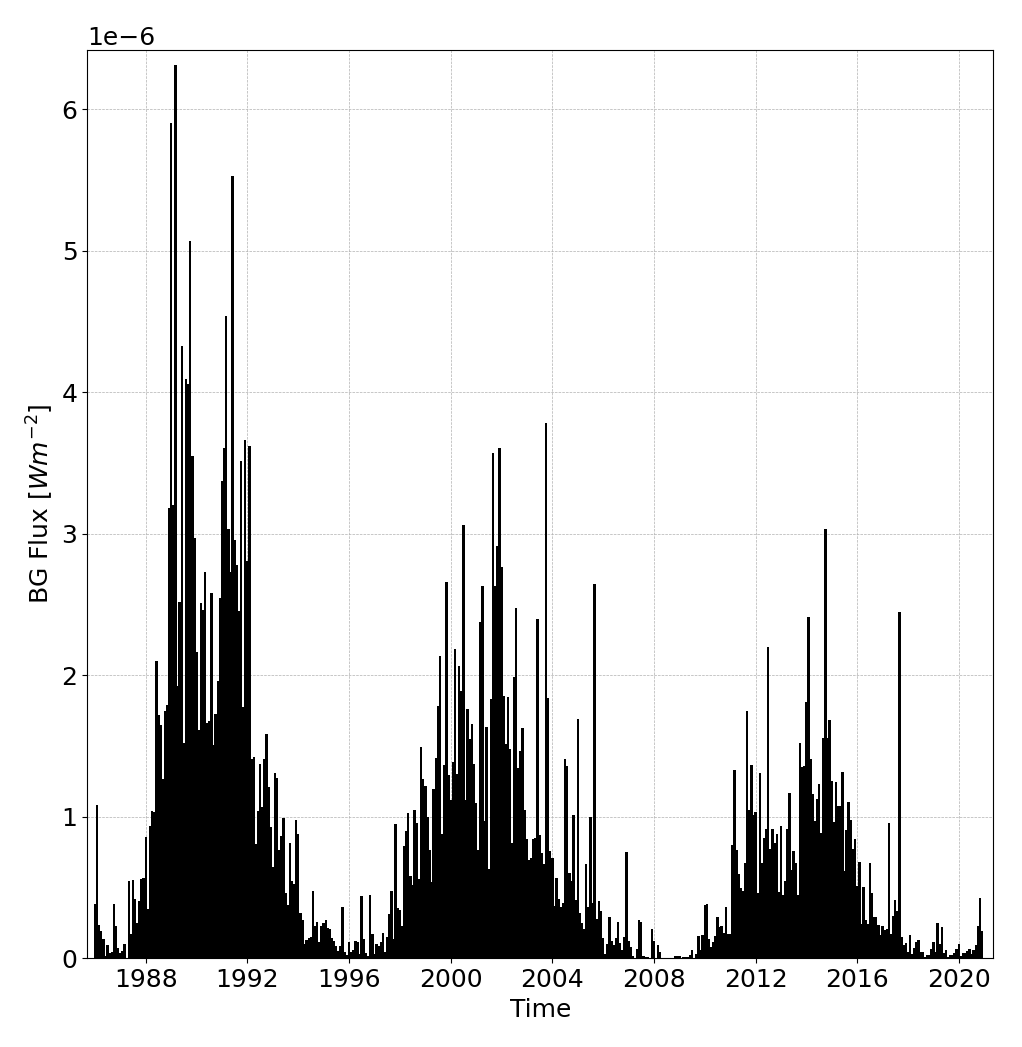}
\caption{The plot reports the monthly average of the background flux $f_ {BG}$ calculated for all the events identified by our procedure in the period from 1986 to 2020.}
\label{fig:bg_month}
\end{figure}

In our algorithm the calculation of the flux level background  $f_{BG}$ is computed from the starting minimum $t_i$ preceding the event candidate (maximum). The figure \ref{fig:bg_month} shows the monthly average of the $f_ {BG}$ calculated for all the events identified by our procedure in the period from 1986 to 2020. Its dependence on solar activity and the 11-years cycle is evident.

Since each minimum point is considered as a possible flare starting time $t_i$, a "pre-flare" background flux $f_{BG}$, obtained as the mean flux in a time interval $[t_i - 60s,t_i]$, is calculated and associated to each starting time $t_i$. 
Once the value of $f_{BG}$ has been defined, the flux $f_{end}$ associated with the end point of the event is defined by the relation: 
\begin{equation}
\label{eq:endpoint}
    f_{end} \leq f_{BG}+f_{noise}
\end{equation}
where $f_{noise}=2\cdot10^{-8}Wm\textsuperscript{-2}$ is the noise level for the typical GOES fluxes and is calculated during quiescent time periods \citep{aschwanden_algo}.

When the condition described by the relationship \ref{eq:endpoint} is not satisfied within the analyzed 24-hour batch, it means that the decay phase of the flare has not finished within the analyzed interval. In this case we assume the end of the analyzed batch as the end of the event. The effects of this simplification, which will be taken into account in a new version of the algorithm, were discussed at the beginning of the section.

At the end of these operations a list of candidate flaring intervals is produced. The procedure iterates through the list and looks for the maximum flux $f_{max}$ within the considered interval. The condition for the definition of $f_{max}$ is:
\begin{equation}
\label{eq:criterio}
    f_{max}>f_{BG} + f_{threshold}
\end{equation}
where the background flux $f_{BG}$ is referred to the starting minimum and $f_{threshold}$ is an adjustable parameter.
The term $f_{BG}$ in condition \ref{eq:criterio} accounts for the soft X-ray flux level background and guarantees that only events with a positive absolute intensity $I_F=f_{max}-f_{BG}>0$ are considered.
It is important to underline that this definition allows to identify class A and B flares even during the phases of maximum solar activity.

\begin{figure}
\centering
\includegraphics[width=0.4\textwidth]{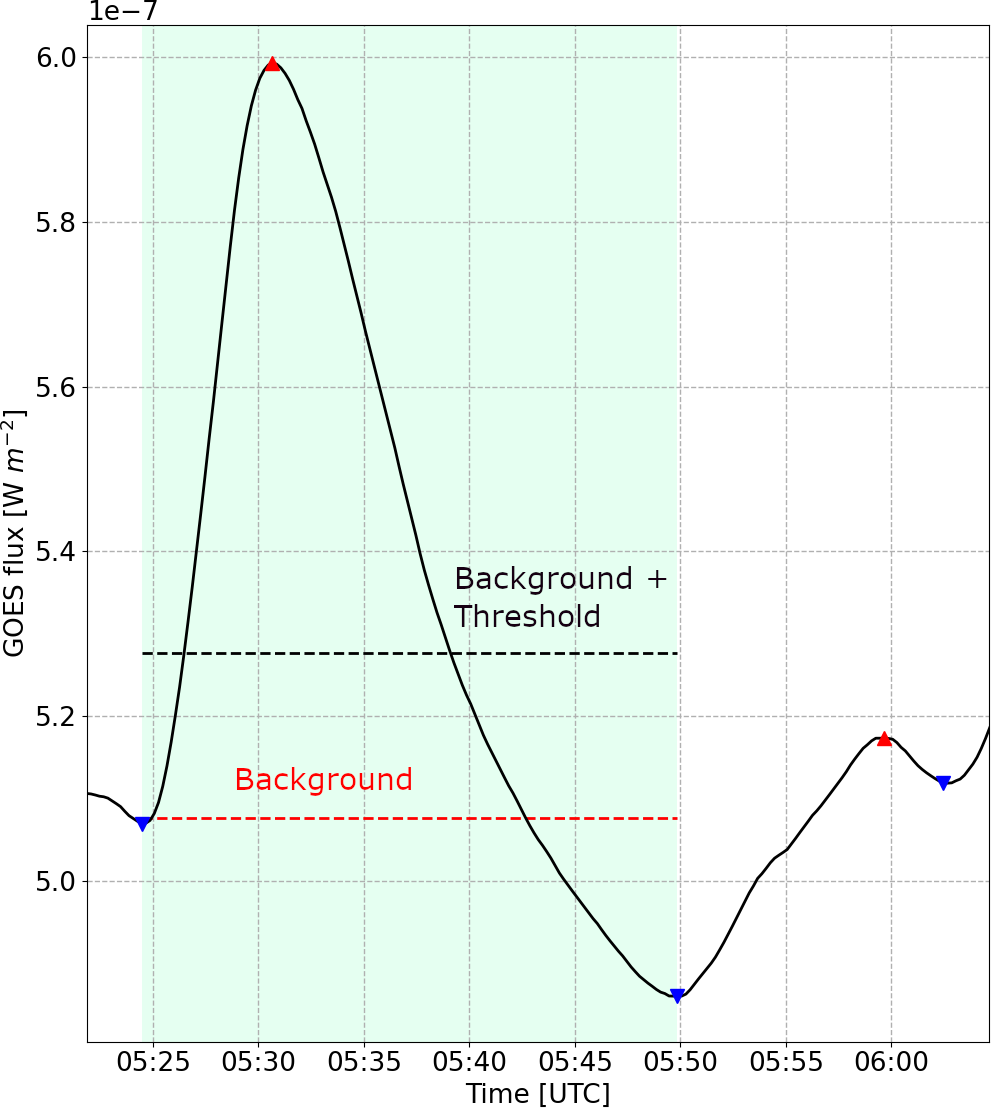}
\caption{Detail about the definition of flare event. The background value $f_{BG}$ (red dashed line) is computed from the starting minimum $t_i$ (first blue downward triangle). The threshold value $f_{threshold}$ is added to background value and the ending minimum $f_{end}$ is chosen (second blue downward triangle). Since the maximum (first red upward triangle) meets Condition \ref{eq:criterio}, a flare is detected in the highlighted area. This figure has been obtained using $f_{threshold}=f_{noise}=2 \cdot 10^{-8} Wm\textsuperscript{-2}$. Data refer to 1\textsuperscript{st} January 2001.}
\label{fig:detection}
\end{figure}

As for the choice of the $f_{threshold}$ parameter, we set $f_{threshold}=f_{noise}$. The $f_{threshold}$ parameter influences the identification of events. As its value increases, the procedure selects more energetic events, on the contrary by decreasing the value too much we risk confusing stochastic fluctuations of the signal for low energy events. 
Our choice, i.e. $f_{threshold}=f_{noise}$, reduces the probability that fluctuations produced by noise are confused with flares and places a limit on the intensity of the identifiable flares which is equal to about an A2.0 class. 
The Figure \ref{fig:detection} shows an example of flare identification. The main thresholds used by the algorithm are also shown.

\begin{figure}
\centering
\includegraphics[width=0.4\textwidth]{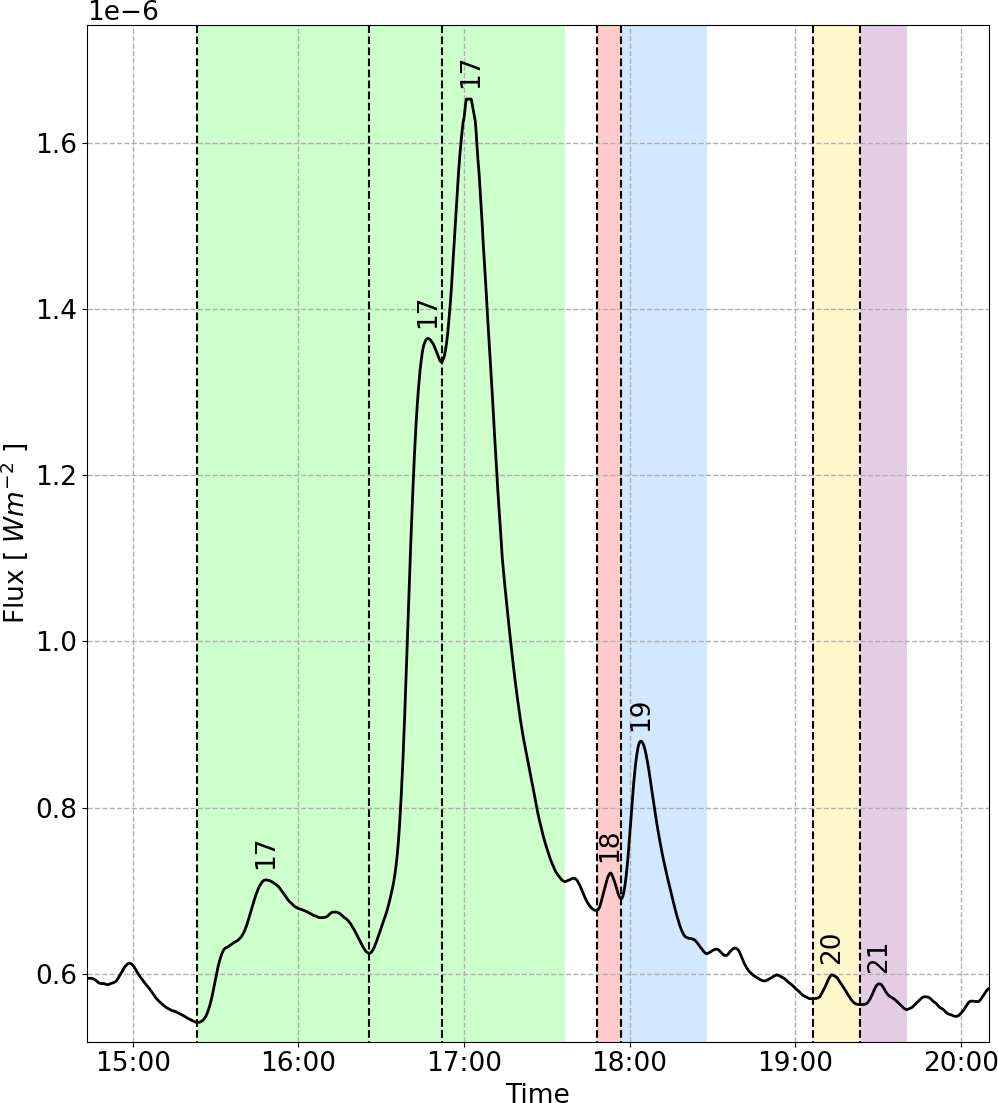}
\caption{Dashed vertical lines represent flare starting times, while highlighted areas refer to the duration of each event. Different colors have been used to group flares with the same label.
The first two flares in the green time interval do not reach an ending minimum point that satisfies Condition \ref{eq:endpoint}, therefore they are associated with the last flare of the group, with the same \textit{Multiple ID}.
Conversely, the ending point of flare 18, even if representing the starting point of flare 19, meets Condition \ref{eq:endpoint}, so these two flares are associated with two different  \textit{Multiple ID}. Data refer to 1\textsuperscript{st} January 2001.}
\label{fig:consecutive}
\end{figure}

\subsection{Overlapping flare events}

The final list can contain flares with overlapping time intervals because the same ending minimum (flare endpoint) can be matched by more than one starting point.
Every time that a flare has not yet reached its ending point before another flare starts, the duration of the first flare is truncated at the beginning of the second one.
\cite{aschwanden_algo} already pointed out how flare duration can be underestimated by this truncation process.
To solve this problem we introduce a system of labels which can preserve the information about the original ending point. Every time that the duration of the first flare is truncated at the beginning of the second one,  the two (or more) events are labelled with the same \textit{Multiple ID}. 
In this way, while performing data analysis, these events can be considered both individually or as multiple events. The total duration of the multiple event can be obtained using the group ID and considering the starting point of the first flare of the group and the end point of the last event of the group.
This labeling system can add a layer of complexity and open new possibilities for the analysis of flare characteristics. 
Figure \ref{fig:consecutive} shows how the time truncation and labelling procedures work.

\subsection{Event attributes and descriptions}
The catalogue consist of records of 10 fields, each of which represented an attribute of the identified event. Table \ref{table:1} reports a list of the event attributes and a short description of the attribute.\\
\begin{table}[h!]
\centering
\renewcommand*{\arraystretch}{1.4}

\begin{tabular}{ |p{6cm}|p{9cm}|  }
\hline
\multicolumn{2}{|c|}{Event attributes and descriptions} \\
\hline
Attribute & Description \\
\hline
Event ID & Unique ID to identify single flare events  \\

Start time [UTC] & Estimated event start time    \\

End time [UTC] & Estimated event end time  \\
Peak time [UTC] & Estimated event maximum (peak) flux time \\
Peak Flux [Wm\textsuperscript{-2}] & As in the GOES catalog without subtracting the associated background flux  \\
Flare class & As in the GOES catalog, i.e., A, B, C, M, X   \\
Background Flux [Wm\textsuperscript{-2}] & The evaluated background level ($f_{BG}$) of the GOES X-ray emission close to the revealed event    \\
Multiple ID & A unique ID for overlapping flares. This ID can be used to group flaring events and test how flare characteristics change when grouping the flare events. In principle, the same Multiple ID does not mean that the flares show a homologous behavior but simply that their signal is identified overlapping a previous event and most likely they come from the same active region \\
Total flux [Wm\textsuperscript{-2}] & Integral of the x-ray flux during the flare, as a further estimate of the total energy of an event \\
\hline
\end{tabular}
\caption{Event attributes and short description of the various attributes.}
\label{table:1}
\end{table}

\section{XRS flare detection and catalogue population}

Our catalogue lists 334123 events, identified in the GOES XRS data in period from 1986 to 2020, used to populate our catalogue. The GOES catalogue reports 61375 events for the same period.
The first check we carried out is a comparison between the entries of the official GOES catalogue and those produced in output by our procedure. GOES catalogue data have been retrieved using the Python module \textit{sunpy} \citep{mumford2015sunpy}. Since our list has been produced using the GOES SXR flux data scaled appropriately to obtain true fluxes, while the official GOES catalogue, currently available, does not implement this correction, before carrying out the comparison of the events we have appropriately rescaled the SXR flux data of GOES. Therefore, we repeated the scaling procedure mentioned in the GOES SXR Operational Data document on the data of the GOES catalogue. After this operation we have produced two lists of events with the same scaling and therefore correctly scaled between them for a final comparison.
 
\subsection{Comparison with GOES catalogue}
\begin{figure*}
\centering
\includegraphics[width=0.9\textwidth]{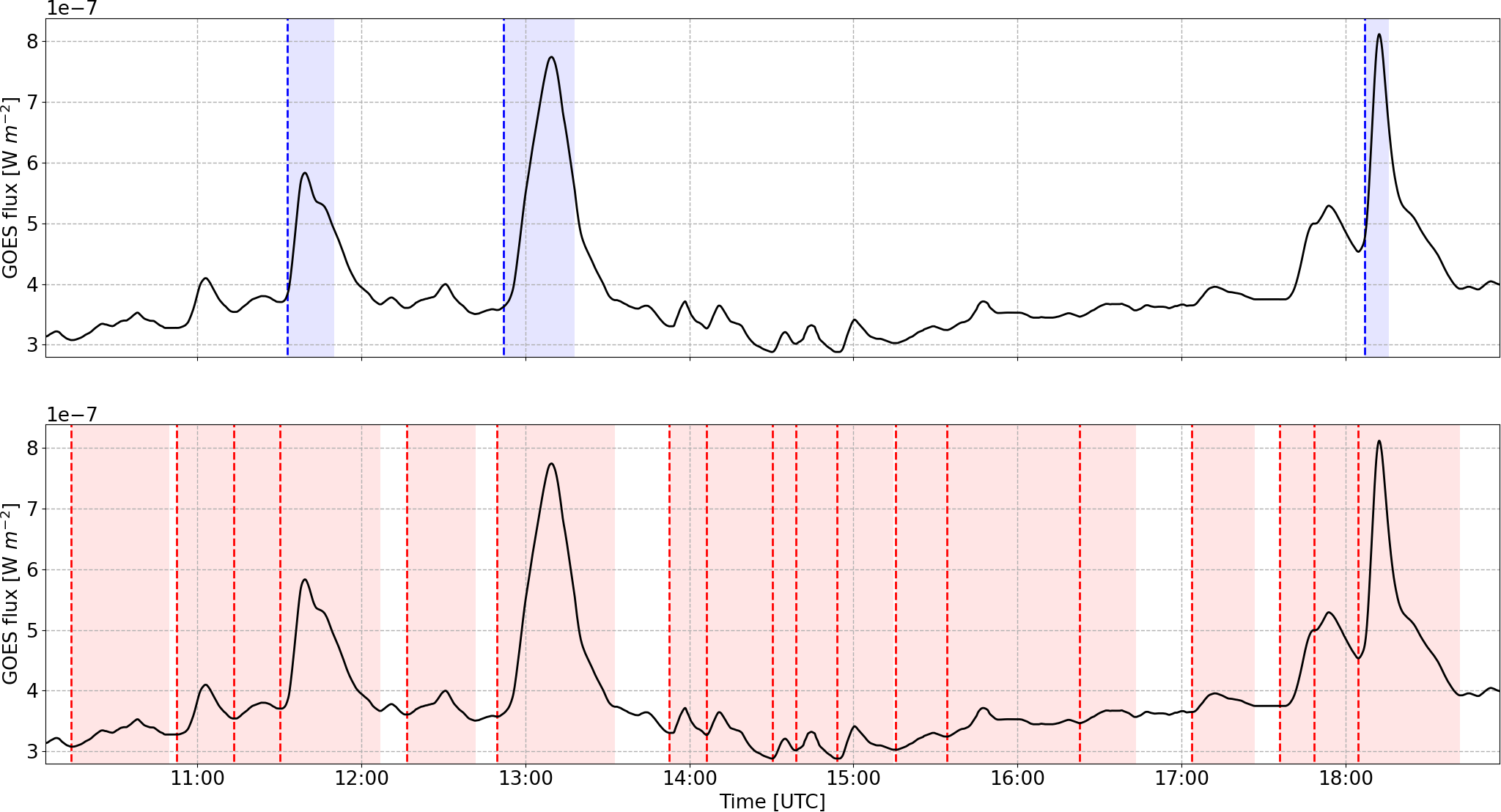}
\caption{Comparison between the two flare catalogs. Dashed lines are used to mark the starting points of events, while the highlighted areas represent the duration of the event. The upper panel, marked with blue, refers to the GOES catalog, while the lower panel and the red color refer to the catalog produced in this study. Data refer to 3\textsuperscript{rd} March 2001.}
\label{fig:comparison}
\end{figure*}

\begin{figure}
\centering
\includegraphics[width=0.4\textwidth]{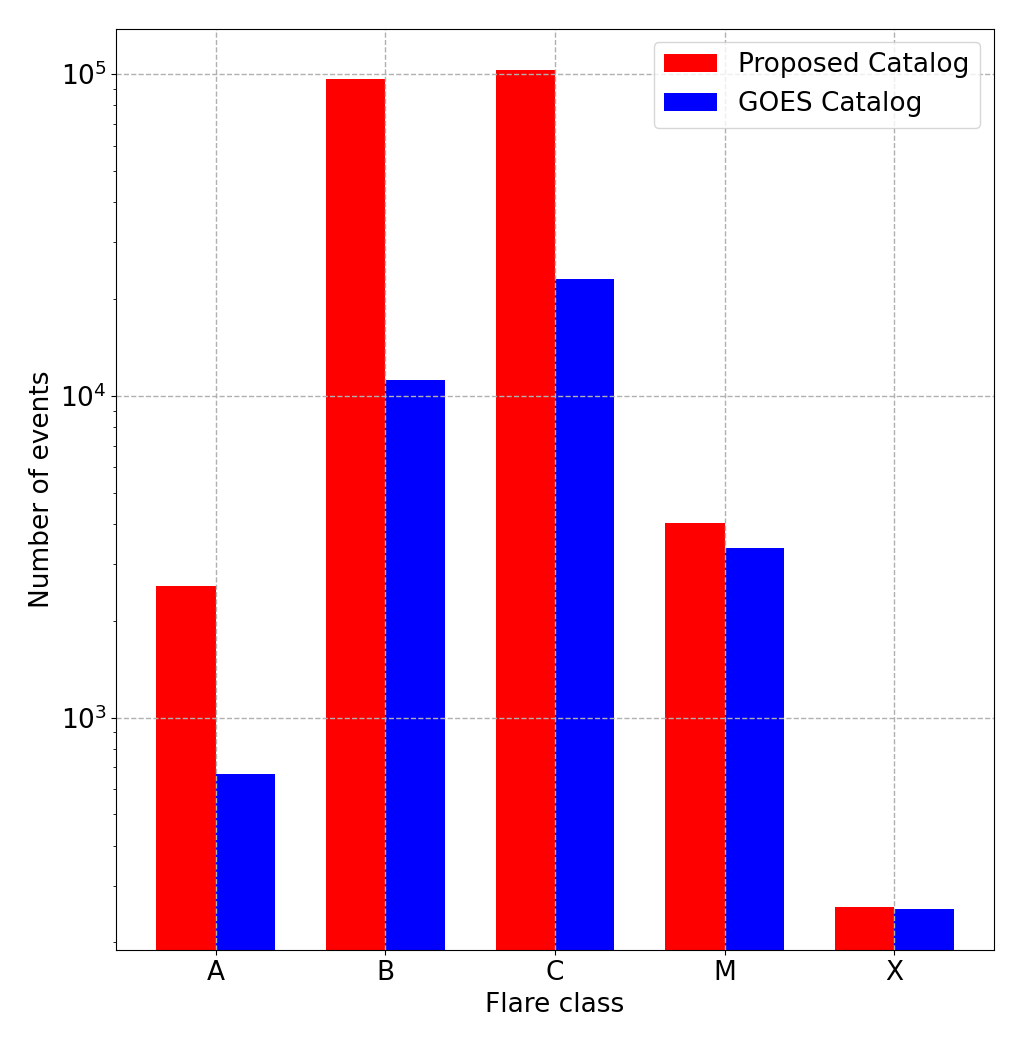}
\caption{Histogram of the number of flares by classes for the period from 1998 to 2020.
The catalogue produced by our procedure increases the statistics of the events of all classes with respect to the events listed in the GOES catalogue. The increase of events in the different classes can be appreciated by observing the number of events listed in our catalogue (red bars) compared to those associated with the GOES catalogue (blue bars).}
\label{fig:nflares}
\end{figure}

We performed various types of comparisons between the two catalogs, both qualitative and quantitative. First of all, we made several qualitative comparisons between the two lists. We compared the light curves and flares listed by GOES soft-X data and those identified by our procedure. In Figure \ref{fig:comparison} we show a typical example of these comparisons. It is evident how our procedure identifies a greater number of flares compared to those listed by GOES.\\
To quantify the difference in identified flares we classified the events in the usual way and we obtained the histogram shown in Figure\ref{fig:nflares}. The comparison is limited to the events identified in the period 1998-2020. This because before May 1997 \citep{swalwell2018} GOES flare list does not show the same event timing definition used in the following period. In the GOES list soft X-ray profiles are used to define the flare event duration only after May 1997, while previous listed events make use of $H\alpha$ data to obtain start and end times \citep{Veronig2002}. In this period the proposed list contains 205989 flares, against the 38517 listed by GOES catalog.
Although we do not believe that this difference has a real impact on the determination of intensities, but only in the timing of events, we prefer to restrict the analyzes to a homogeneous data period. 

\subsection{Peak flux distributions}
As mentioned in Section \ref{section:introduction} GOES catalog has already been used to analyse the distribution of solar flare peak fluxes. We compare this distribution for the GOES catalog with our list of flares for the period from 1998 to 2020.
\begin{figure*}[ht]
\centering
\includegraphics[width=0.67\textwidth]{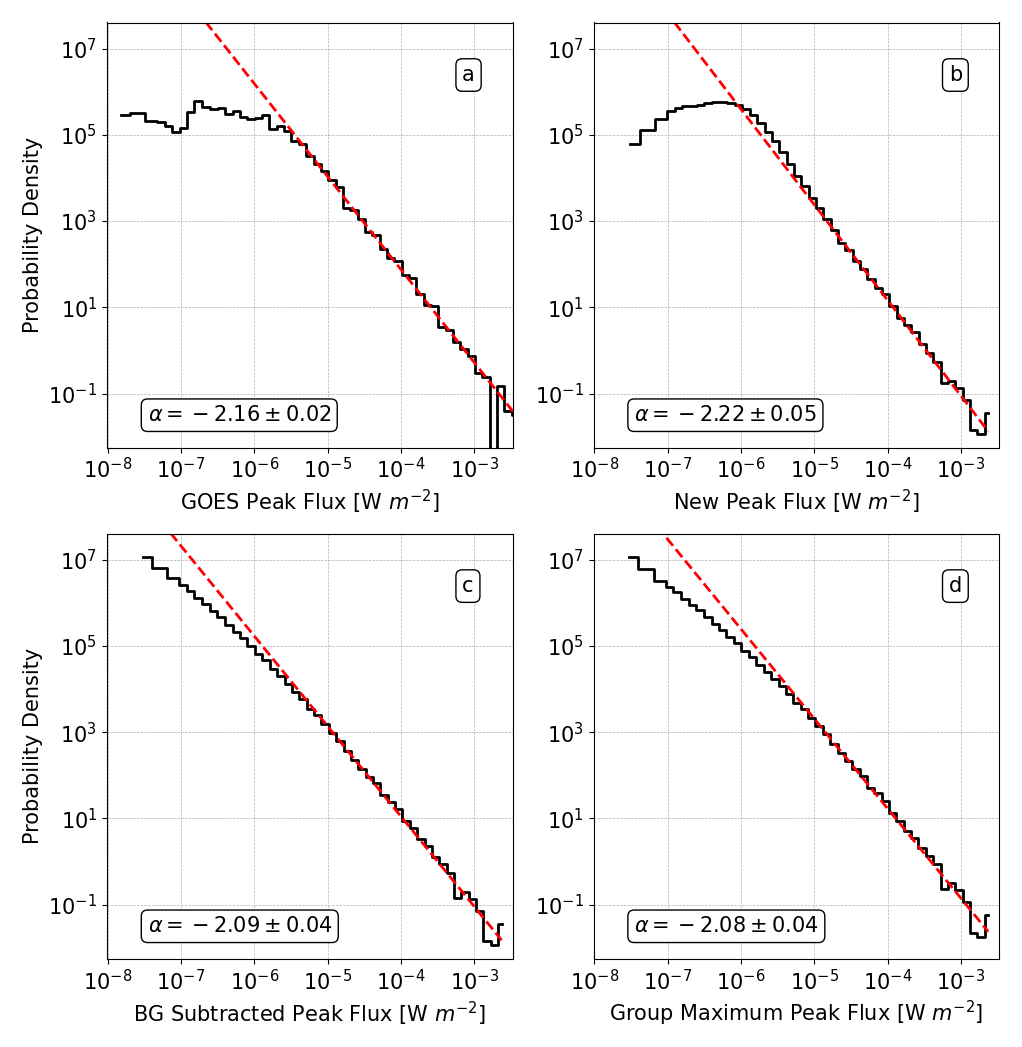}
\caption{
In this figure we use GOES list and the proposed one  to obtain peak flux probability density functions for the period 1998-2020. Each distribution is obtained with a different criterium: \textit{Panel a)} peak fluxes listed in the GOES catalog; \textit{Panel b)} peak fluxes listed in the catalog proposed in this study; \textit{Panel c)} and \textit{Panel d)}  show peak fluxes obtained from our catalog once the $f_{BG}$ is subtracted. \textit{Panel c)}  is obtained considering all 205989 events listed in our catalog, while \textit{Panel d)}  is computed using the most intense event flux within a sequence recognized as multiple (130925 events).
}
\label{fig:distributions}
\end{figure*}

To make this comparison we treat the flares peak flux as a random variable $X$ and consider the Probability Density Function (PDF) $p(x)$, which is related to the probability of obtaining a flux value between $F_{min}$ and $F_{max}$ through the formula:

\begin{equation}
    Pr(F_{min}\leq X \leq F_{max})= \int_{F_{min}}^{F_{max}} p(x)dx
\end{equation}

\noindent The expected shape for a power law PDF is the following:
\begin{equation}\label{eq:pdf}
    p(x)= \frac{\alpha -1 }{x_{min}} (\frac{x_{min}}{x})^{\alpha}
\end{equation}
with $\alpha$ index of the power-law and $x_{min}$ the minimum value of flux for which the power-law holds. The PDF obtained using GOES list and the one obtained using our list can be seen in Figure \ref{fig:distributions} (\textit{Panel a} and \textit{b}). The list of events obtained with our algorithm is missing flares below the A2.0 class (with a peak flux lower than $f_{noise}$) because of the built-in Condition \ref{eq:criterio}. 

In Figure \ref{fig:distributions} different distributions of peak fluxes are shown both from the GOES catalogue and from ours. Relevant are panels c) and d) which show our results and take into account the background flux contribution. We hypothesized power law distributions (red lines) and validated the hypothesis by means of the Kolmogorov-Smirnov (K-S test) statistics. This test gives us a measure of the goodness-of-fit of the data \citep{goldstein2004problems}.
We used the Python package \textit{powerlaw} \citep{alstott2014powerlaw} to perform the K-S test and estimate the power law region, including its lower ($x_{min}$) and upper ($x_{max}$) bounds (which must hold more decades to have statistical significance).
This package returns the bound values and the value of power law index $\alpha$ by means of a linear fit.

In more detail, the results obtained can be listed as follows:
\begin{itemize}
\item GOES list (without correction for $f_{BG}$): we find a power law index $\alpha=-2.16\pm0.02$ and $x_{min}=7.4\cdot10^{-6}$ Wm\textsuperscript{-2}. The fit (red line) is shown in Figure \ref{fig:distributions} \textit{Panel a)}.
\item Our catalog (without correction for $f_{BG}$): we find a power law index $\alpha=-2.22\pm0.05$ and $x_{min}=1.8\cdot10^{-5}$ Wm\textsuperscript{-2}. The fit (red line) is shown in Figure\ref{fig:distributions} \textit{Panel b)}.
\item Our catalog ($f_ {BG}$ subtracted from peak fluxes): we find a power law index $\alpha= -2.09\pm0.04$ and $x_{min}=5.5\cdot10^{-6}$ Wm\textsuperscript{-2}. The fit (red line) is shown in Figure\ref{fig:distributions} \textit{Panel c)}.
\item Our catalog ($f_{BG}$ subtracted from peak fluxes), we only show the maximum flux in the flare set with the same multiplicity flag: we find a power law index $\alpha=-2.08\pm0.04$ and $x_{min}=5.9\cdot10^{-6}$ Wm \textsuperscript{-2}. The fit (red line) is shown in Figure\ref{fig:distributions} \textit{Panel d)}.
\end{itemize}
The results found demonstrate that our power-law indices, calculated taking into account the $f_ {BG}$, agree with the values reported in the literature (see the brief discussion in the next section). For completeness, we underline that the algorithm used assumes that there is no upper bound to the empirical power-law distribution.

\section{Conclusions and access to the catalogue}
In this paper we present a catalogue of SXR events with a whole coverage over about three solar cycles based on GOES XRS data. A total of 334123 events have been identified and analyzed. If we limit the analysis to the period 1998-2020, taking into account that NOAA did not use a standard definition for flare timing prior to May 1997, we then identify 205989 events, compared to the 38517 events listed in the official GOES catalog during the same period. In both cases we have an increase of about a factor 5 in the number of identified events. Obviously not all classes are incremented by the same factor. But the key point is that the statistics of the sample to be analyzed is greatly increased.\\
This result allows us to carry out a preliminary statistical analysis on our data. An analysis was performed on the 1998-2020 events in order to study the distributions of peak fluxes under different conditions. 
In particular, we carried out both a direct comparison of the distributions of the peak fluxes between our catalog and GOES list (see Figure \ref{fig:distributions} \textit{Panels a, b}), and a comparison when we subtract from the peak flux the estimated background level of the GOES X-ray emission associated to the identified event.

If we limit the analysis to the latter case, in which we take into account the estimate of the background level, we observe (see Figure \ref{fig:distributions}, \textit{Panels c, d}) that the power law index is substantially the same if we consider all the 205989 identified events $(\alpha=-2.09\pm0.04)$ or we consider only the most intense events (130925 events) in the various sequences of overlapping events $(\alpha=-2.08\pm0.04)$. But above all it must be emphasized that distributions of all the 205989 identified events show a clear power-law behavior over more than three decades with a power law index $\alpha=-2.09\pm0.04$.
The value of $\alpha$ we found is compatible within the errors with the power-law slope of $\alpha_{F}= -1.98\pm0.11$ for the (background-subtracted) soft X-ray peak fluxes distribution reported by \cite{aschwanden_algo}.

This preliminary analysis and the comparison with similar works reported in the literature \citep[e.g.][]{Veronig2002,Yashiro2006,aschwanden_algo} makes us confident of the goodness of the algorithm developed. The next step will consist in using the other physical parameters calculated for the events in order to study other statistical properties of the flares such as the estimated total energy distribution, waiting times, duration or these properties associated with the multiplicity of events. These analyses, which have already begun, will be the subject of future work. 

The whole list of identified flares and all the outputs of the procedure are available through the web page 
\url{https://github.com/nplutino/FlareList} together with Python scripts to retrieve selected data that match user-defined criteria. 

\section{Acknowledgments}
We aknowledge the use of data from GOES satellites, concerning solar signal in the soft-X range (\url{https://www.ngdc.noaa.gov/stp/satellite/goes/dataaccess.html}) and the use of NOAA SWPC flare lists (\url{https://www.ngdc.noaa.gov/stp/solar/solarflares.html}). This work is partially supported by the University of Rome Tor Vergata  "Beyond Borders" grant (D.R. 1347 of 29 May 2019). The authors thank two anonymous reviewers for their significant comments that helped improve the manuscript.

\bibliographystyle{model5-names}
\biboptions{authoryear}
\bibliography{refs}

\begin{thebibliography}{43}
\expandafter\ifx\csname natexlab\endcsname\relax\def\natexlab#1{#1}\fi
\providecommand{\url}[1]{\texttt{#1}}
\providecommand{\href}[2]{#2}
\providecommand{\path}[1]{#1}
\providecommand{\DOIprefix}{doi:}
\providecommand{\ArXivprefix}{arXiv:}
\providecommand{\URLprefix}{URL: }
\providecommand{\Pubmedprefix}{pmid:}
\providecommand{\doi}[1]{\href{http://dx.doi.org/#1}{\path{#1}}}
\providecommand{\Pubmed}[1]{\href{pmid:#1}{\path{#1}}}
\providecommand{\bibinfo}[2]{#2}
\ifx\xfnm\relax \def\xfnm[#1]{\unskip,\space#1}\fi
\bibitem[{Alstott \& Bullmore(2014)}]{alstott2014powerlaw}
\bibinfo{author}{Alstott, J.}, \& \bibinfo{author}{Bullmore, D.~P.}
  (\bibinfo{year}{2014}).
\newblock \bibinfo{title}{powerlaw: a python package for analysis of
  heavy-tailed distributions}.
\newblock {\it \bibinfo{journal}{PloS one}\/},  {\it
  \bibinfo{volume}{9}\/}\bibinfo{issue}{(1)}.
\bibitem[{Aschwanden \& Freeland(2012)}]{aschwanden_algo}
\bibinfo{author}{Aschwanden, M.~J.}, \& \bibinfo{author}{Freeland, S.~L.}
  (\bibinfo{year}{2012}).
\newblock \bibinfo{title}{Automated solar flare statistics in soft x-rays over
  37 years of goes observations: The invariance of self-organized criticality
  during three solar cycles}.
\newblock {\it \bibinfo{journal}{The Astrophysical Journal}\/},  {\it
  \bibinfo{volume}{754}\/}\bibinfo{issue}{(2)}, \bibinfo{pages}{112}.
\bibitem[{{Berrilli} et~al.(2019){Berrilli}, {Casolino}, {Cristaldi}, {Del
  Moro}, {Forte}, {Giovannelli}, {Martucci}, {Merg{\'e}}, {Napoletano},
  {Narici}, {Pietropaolo}, {Pucacco}, {Rizzo}, {Scardigli} \&
  {Sparvoli}}]{Berrilli2019}
\bibinfo{author}{{Berrilli}, F.}, \bibinfo{author}{{Casolino}, M.},
  \bibinfo{author}{{Cristaldi}, A.}, \bibinfo{author}{{Del Moro}, D.},
  \bibinfo{author}{{Forte}, R.}, \bibinfo{author}{{Giovannelli}, L.},
  \bibinfo{author}{{Martucci}, M.}, \bibinfo{author}{{Merg{\'e}}, M.},
  \bibinfo{author}{{Napoletano}, G.}, \bibinfo{author}{{Narici}, L.},
  \bibinfo{author}{{Pietropaolo}, E.}, \bibinfo{author}{{Pucacco}, G.},
  \bibinfo{author}{{Rizzo}, A.}, \bibinfo{author}{{Scardigli}, S.}, \&
  \bibinfo{author}{{Sparvoli}, R.} (\bibinfo{year}{2019}).
\newblock \bibinfo{title}{{Introducing SWERTO: A regional space weather
  service}}.
\newblock {\it \bibinfo{journal}{Nuovo Cimento C Geophysics Space Physics
  C}\/},  {\it \bibinfo{volume}{42}\/}\bibinfo{issue}{(1)},
  \bibinfo{pages}{47}. \DOIprefix\doi{10.1393/ncc/i2019-19047-4}.
\bibitem[{{Berrilli} \& {Giovannelli}(2022)}]{Berrilli2022}
\bibinfo{author}{{Berrilli}, F.}, \& \bibinfo{author}{{Giovannelli}, L.}
  (\bibinfo{year}{2022}).
\newblock \bibinfo{title}{{The Great Aurora of 4 February 1872 observed by
  Angelo Secchi in Rome}}.
\newblock {\it \bibinfo{journal}{Journal of Space Weather and Space
  Climate}\/},  {\it \bibinfo{volume}{12}\/}, \bibinfo{pages}{3}.
  \DOIprefix\doi{10.1051/swsc/2021046}.
  \href{http://arxiv.org/abs/2201.01171}{\tt arXiv:2201.01171}.
\bibitem[{{Bobrinsky} \& {Del Monte}(2010)}]{esa_ssa2010}
\bibinfo{author}{{Bobrinsky}, N.}, \& \bibinfo{author}{{Del Monte}, L.}
  (\bibinfo{year}{2010}).
\newblock \bibinfo{title}{{The space situational awareness program of the
  European Space Agency}}.
\newblock {\it \bibinfo{journal}{Cosmic Research}\/},  {\it
  \bibinfo{volume}{48}\/}\bibinfo{issue}{(5)}, \bibinfo{pages}{392--398}.
  \DOIprefix\doi{10.1134/S0010952510050035}.
\bibitem[{Boffetta et~al.(1999)Boffetta, Carbone, Giuliani, Veltri \&
  Vulpiani}]{Boffetta1999}
\bibinfo{author}{Boffetta, G.}, \bibinfo{author}{Carbone, V.},
  \bibinfo{author}{Giuliani, P.}, \bibinfo{author}{Veltri, P.}, \&
  \bibinfo{author}{Vulpiani, A.} (\bibinfo{year}{1999}).
\newblock \bibinfo{title}{Power laws in solar flares: self-organized
  criticality or turbulence?}
\newblock {\it \bibinfo{journal}{Physical review letters}\/},  {\it
  \bibinfo{volume}{83}\/}\bibinfo{issue}{(22)}, \bibinfo{pages}{4662}.
\bibitem[{Charbonneau et~al.(2001)Charbonneau, McIntosh, Liu \&
  Bogdan}]{Charbonneau2001}
\bibinfo{author}{Charbonneau, P.}, \bibinfo{author}{McIntosh, S.~W.},
  \bibinfo{author}{Liu, H.-L.}, \& \bibinfo{author}{Bogdan, T.~J.}
  (\bibinfo{year}{2001}).
\newblock \bibinfo{title}{Avalanche models for solar flares (invited review)}.
\newblock {\it \bibinfo{journal}{Solar Physics}\/},  {\it
  \bibinfo{volume}{203}\/}\bibinfo{issue}{(2)}, \bibinfo{pages}{321--353}.
\bibitem[{Cicogna et~al.(2021)Cicogna, Berrilli, Calchetti, Del~Moro,
  Giovannelli, Benvenuto, Campi, Guastavino \& Piana}]{cicogna2021flare}
\bibinfo{author}{Cicogna, D.}, \bibinfo{author}{Berrilli, F.},
  \bibinfo{author}{Calchetti, D.}, \bibinfo{author}{Del~Moro, D.},
  \bibinfo{author}{Giovannelli, L.}, \bibinfo{author}{Benvenuto, F.},
  \bibinfo{author}{Campi, C.}, \bibinfo{author}{Guastavino, S.}, \&
  \bibinfo{author}{Piana, M.} (\bibinfo{year}{2021}).
\newblock \bibinfo{title}{Flare-forecasting algorithms based on high-gradient
  polarity inversion lines in active regions}.
\newblock {\it \bibinfo{journal}{The Astrophysical Journal}\/},  {\it
  \bibinfo{volume}{915}\/}\bibinfo{issue}{(1)}, \bibinfo{pages}{38}.
\bibitem[{Crosby et~al.(1993)Crosby, Aschwanden \& Dennis}]{crosby1993}
\bibinfo{author}{Crosby, N.~B.}, \bibinfo{author}{Aschwanden, M.~J.}, \&
  \bibinfo{author}{Dennis, B.~R.} (\bibinfo{year}{1993}).
\newblock \bibinfo{title}{Frequency distributions and correlations of solar
  x-ray flare parameters}.
\newblock {\it \bibinfo{journal}{Solar Physics}\/},  {\it
  \bibinfo{volume}{143}\/}\bibinfo{issue}{(2)}, \bibinfo{pages}{275--299}.
\bibitem[{{Crown}(2012)}]{Crown2012}
\bibinfo{author}{{Crown}, M.~D.} (\bibinfo{year}{2012}).
\newblock \bibinfo{title}{{Validation of the NOAA Space Weather Prediction
  Center's solar flare forecasting look-up table and forecaster-issued
  probabilities}}.
\newblock {\it \bibinfo{journal}{Space Weather}\/},  {\it
  \bibinfo{volume}{10}\/}, \bibinfo{pages}{S06006}.
  \DOIprefix\doi{10.1029/2011SW000760}.
\bibitem[{Fletcher(2005)}]{Fletcher2005}
\bibinfo{author}{Fletcher, L.} (\bibinfo{year}{2005}).
\newblock \bibinfo{title}{The observational motivation for computational
  advances in solar flare physics}.
\newblock {\it \bibinfo{journal}{Space science reviews}\/},  {\it
  \bibinfo{volume}{121}\/}\bibinfo{issue}{(1)}, \bibinfo{pages}{141--152}.
\bibitem[{{Giuliani} et~al.(2000){Giuliani}, {Carbone}, {Veltri}, {Boffetta} \&
  {Vulpiani}}]{Giuliani2000}
\bibinfo{author}{{Giuliani}, P.}, \bibinfo{author}{{Carbone}, V.},
  \bibinfo{author}{{Veltri}, P.}, \bibinfo{author}{{Boffetta}, G.}, \&
  \bibinfo{author}{{Vulpiani}, A.} (\bibinfo{year}{2000}).
\newblock \bibinfo{title}{{Waiting time statistics in solar flares}}.
\newblock {\it \bibinfo{journal}{Physica A Statistical Mechanics and its
  Applications}\/},  {\it \bibinfo{volume}{280}\/}, \bibinfo{pages}{75--80}.
  \DOIprefix\doi{10.1016/S0378-4371(99)00620-2}.
\bibitem[{Goldstein et~al.(2004)Goldstein, Morris \&
  Yen}]{goldstein2004problems}
\bibinfo{author}{Goldstein, M.~L.}, \bibinfo{author}{Morris, S.~A.}, \&
  \bibinfo{author}{Yen, G.~G.} (\bibinfo{year}{2004}).
\newblock \bibinfo{title}{Problems with fitting to the power-law distribution}.
\newblock {\it \bibinfo{journal}{The European Physical Journal B-Condensed
  Matter and Complex Systems}\/},  {\it
  \bibinfo{volume}{41}\/}\bibinfo{issue}{(2)}, \bibinfo{pages}{255--258}.
\bibitem[{{Ishii}(2017)}]{Ishii2017}
\bibinfo{author}{{Ishii}, M.} (\bibinfo{year}{2017}).
\newblock \bibinfo{title}{{Japanese space weather research activities}}.
\newblock {\it \bibinfo{journal}{Space Weather}\/},  {\it
  \bibinfo{volume}{15}\/}\bibinfo{issue}{(1)}, \bibinfo{pages}{26--35}.
  \DOIprefix\doi{10.1002/2016SW001531}.
\bibitem[{{Kashapova} et~al.(2021){Kashapova}, {Broomhall}, {Larionova},
  {Kupriyanova} \& {Motyk}}]{Kashapova2021}
\bibinfo{author}{{Kashapova}, L.~K.}, \bibinfo{author}{{Broomhall}, A.-M.},
  \bibinfo{author}{{Larionova}, A.~I.}, \bibinfo{author}{{Kupriyanova}, E.~G.},
  \& \bibinfo{author}{{Motyk}, I.~D.} (\bibinfo{year}{2021}).
\newblock \bibinfo{title}{{The morphology of average solar flare time profiles
  from observations of the Sun's lower atmosphere}}.
\newblock {\it \bibinfo{journal}{Monthly Notices of the Royal Astronomical
  Society}\/},  {\it \bibinfo{volume}{502}\/}\bibinfo{issue}{(3)},
  \bibinfo{pages}{3922--3931}. \DOIprefix\doi{10.1093/mnras/stab276}.
  \href{http://arxiv.org/abs/2102.02596}{\tt arXiv:2102.02596}.
\bibitem[{Li et~al.(2016)Li, Feng, Zhang, Liu \& Gan}]{li2016}
\bibinfo{author}{Li, Y.-P.}, \bibinfo{author}{Feng, L.},
  \bibinfo{author}{Zhang, P.}, \bibinfo{author}{Liu, S.-M.}, \&
  \bibinfo{author}{Gan, W.-Q.} (\bibinfo{year}{2016}).
\newblock \bibinfo{title}{On the power-law distributions of x-ray fluxes from
  solar flares observed with goes}.
\newblock {\it \bibinfo{journal}{Research in Astronomy and Astrophysics}\/},
  {\it \bibinfo{volume}{16}\/}\bibinfo{issue}{(10)}, \bibinfo{pages}{161}.
\bibitem[{Lin \& Hudson(1976)}]{lin1976non}
\bibinfo{author}{Lin, R.}, \& \bibinfo{author}{Hudson, H.}
  (\bibinfo{year}{1976}).
\newblock \bibinfo{title}{Non-thermal processes in large solar flares}.
\newblock {\it \bibinfo{journal}{Solar Physics}\/},  {\it
  \bibinfo{volume}{50}\/}\bibinfo{issue}{(1)}, \bibinfo{pages}{153--178}.
\bibitem[{Lu et~al.(2021)Lu, Feng, Li, Ying, Li, Gan, Li \& Zhou}]{Lu2021}
\bibinfo{author}{Lu, L.}, \bibinfo{author}{Feng, L.}, \bibinfo{author}{Li, D.},
  \bibinfo{author}{Ying, B.}, \bibinfo{author}{Li, H.}, \bibinfo{author}{Gan,
  W.}, \bibinfo{author}{Li, Y.}, \& \bibinfo{author}{Zhou, J.}
  (\bibinfo{year}{2021}).
\newblock \bibinfo{title}{Catalog and statistical examinations of ly$\alpha$
  solar flares from goes/euvs-e measurements}.
\newblock {\it \bibinfo{journal}{The Astrophysical Journal Supplement
  Series}\/},  {\it \bibinfo{volume}{253}\/}\bibinfo{issue}{(1)},
  \bibinfo{pages}{29}.
\bibitem[{McKinney(2011)}]{pandas}
\bibinfo{author}{McKinney, W.} (\bibinfo{year}{2011}).
\newblock \bibinfo{title}{pandas: a foundational python library for data
  analysis and statistics}.
\newblock {\it \bibinfo{journal}{Python for high performance and scientific
  computing}\/},  {\it \bibinfo{volume}{14}\/}\bibinfo{issue}{(9)},
  \bibinfo{pages}{1--9}.
\bibitem[{Mumford et~al.(2015)Mumford, Christe, P{\'e}rez-Su{\'a}rez, Ireland,
  Shih, Inglis, Liedtke, Hewett, Mayer, Hughitt et~al.}]{mumford2015sunpy}
\bibinfo{author}{Mumford, S.~J.}, \bibinfo{author}{Christe, S.},
  \bibinfo{author}{P{\'e}rez-Su{\'a}rez, D.}, \bibinfo{author}{Ireland, J.},
  \bibinfo{author}{Shih, A.~Y.}, \bibinfo{author}{Inglis, A.~R.},
  \bibinfo{author}{Liedtke, S.}, \bibinfo{author}{Hewett, R.~J.},
  \bibinfo{author}{Mayer, F.}, \bibinfo{author}{Hughitt, K.} et~al.
  (\bibinfo{year}{2015}).
\newblock \bibinfo{title}{Sunpy—python for solar physics}.
\newblock {\it \bibinfo{journal}{Computational Science \& Discovery}\/},  {\it
  \bibinfo{volume}{8}\/}\bibinfo{issue}{(1)}, \bibinfo{pages}{014009}.
\bibitem[{{Murray} et~al.(2017){Murray}, {Bingham}, {Sharpe} \&
  {Jackson}}]{Murray2017}
\bibinfo{author}{{Murray}, S.~A.}, \bibinfo{author}{{Bingham}, S.},
  \bibinfo{author}{{Sharpe}, M.}, \& \bibinfo{author}{{Jackson}, D.~R.}
  (\bibinfo{year}{2017}).
\newblock \bibinfo{title}{{Flare forecasting at the Met Office Space Weather
  Operations Centre}}.
\newblock {\it \bibinfo{journal}{Space Weather}\/},  {\it
  \bibinfo{volume}{15}\/}\bibinfo{issue}{(4)}, \bibinfo{pages}{577--588}.
  \DOIprefix\doi{10.1002/2016SW001579}.
  \href{http://arxiv.org/abs/1703.06754}{\tt arXiv:1703.06754}.
\bibitem[{Park et~al.(2020)Park, Leka, Kusano, Andries, Barnes, Bingham,
  Bloomfield, McCloskey, Delouille, Falconer et~al.}]{park2020comparison}
\bibinfo{author}{Park, S.-H.}, \bibinfo{author}{Leka, K.},
  \bibinfo{author}{Kusano, K.}, \bibinfo{author}{Andries, J.},
  \bibinfo{author}{Barnes, G.}, \bibinfo{author}{Bingham, S.},
  \bibinfo{author}{Bloomfield, D.~S.}, \bibinfo{author}{McCloskey, A.~E.},
  \bibinfo{author}{Delouille, V.}, \bibinfo{author}{Falconer, D.} et~al.
  (\bibinfo{year}{2020}).
\newblock \bibinfo{title}{A comparison of flare forecasting methods. iv.
  evaluating consecutive-day forecasting patterns}.
\newblock {\it \bibinfo{journal}{The Astrophysical Journal}\/},  {\it
  \bibinfo{volume}{890}\/}\bibinfo{issue}{(2)}, \bibinfo{pages}{124}.
\bibitem[{{Plainaki} et~al.(2020){Plainaki}, {Antonucci}, {Bemporad},
  {Berrilli}, {Bertucci}, {Castronuovo}, {De Michelis}, {Giardino}, {Iuppa},
  {Laurenza}, {Marcucci}, {Messerotti}, {Narici}, {Negri}, {Nozzoli}, {Orsini},
  {Romano}, {Cavallini}, {Polenta} \& {Ippolito}}]{Plainaki2020}
\bibinfo{author}{{Plainaki}, C.}, \bibinfo{author}{{Antonucci}, M.},
  \bibinfo{author}{{Bemporad}, A.}, \bibinfo{author}{{Berrilli}, F.},
  \bibinfo{author}{{Bertucci}, B.}, \bibinfo{author}{{Castronuovo}, M.},
  \bibinfo{author}{{De Michelis}, P.}, \bibinfo{author}{{Giardino}, M.},
  \bibinfo{author}{{Iuppa}, R.}, \bibinfo{author}{{Laurenza}, M.},
  \bibinfo{author}{{Marcucci}, F.}, \bibinfo{author}{{Messerotti}, M.},
  \bibinfo{author}{{Narici}, L.}, \bibinfo{author}{{Negri}, B.},
  \bibinfo{author}{{Nozzoli}, F.}, \bibinfo{author}{{Orsini}, S.},
  \bibinfo{author}{{Romano}, V.}, \bibinfo{author}{{Cavallini}, E.},
  \bibinfo{author}{{Polenta}, G.}, \& \bibinfo{author}{{Ippolito}, A.}
  (\bibinfo{year}{2020}).
\newblock \bibinfo{title}{{Current state and perspectives of Space Weather
  science in Italy}}.
\newblock {\it \bibinfo{journal}{Journal of Space Weather and Space
  Climate}\/},  {\it \bibinfo{volume}{10}\/}, \bibinfo{pages}{6}.
  \DOIprefix\doi{10.1051/swsc/2020003}.
\bibitem[{Priest \& Forbes(2002)}]{Priest2002}
\bibinfo{author}{Priest, E.}, \& \bibinfo{author}{Forbes, T.}
  (\bibinfo{year}{2002}).
\newblock \bibinfo{title}{The magnetic nature of solar flares}.
\newblock {\it \bibinfo{journal}{The Astronomy and Astrophysics Review}\/},
  {\it \bibinfo{volume}{10}\/}\bibinfo{issue}{(4)}, \bibinfo{pages}{313--377}.
\bibitem[{{Priest}(1986)}]{Priest1981}
\bibinfo{author}{{Priest}, E.~R.} (\bibinfo{year}{1986}).
\newblock \bibinfo{title}{{Magnetohydrodynamic Theories of Solar Flares}}.
\newblock {\it \bibinfo{journal}{Solar Physics}\/},  {\it
  \bibinfo{volume}{104}\/}\bibinfo{issue}{(1)}, \bibinfo{pages}{1--18}.
  \DOIprefix\doi{10.1007/BF00159941}.
\bibitem[{Pucci \& Velli(2013)}]{Pucci2014}
\bibinfo{author}{Pucci, F.}, \& \bibinfo{author}{Velli, M.}
  (\bibinfo{year}{2013}).
\newblock \bibinfo{title}{Reconnection of quasi-singular current sheets: the
  “ideal” tearing mode}.
\newblock {\it \bibinfo{journal}{The Astrophysical Journal Letters}\/},  {\it
  \bibinfo{volume}{780}\/}\bibinfo{issue}{(2)}, \bibinfo{pages}{L19}.
\bibitem[{Reep \& Knizhnik(2019)}]{reep2019determines}
\bibinfo{author}{Reep, J.~W.}, \& \bibinfo{author}{Knizhnik, K.~J.}
  (\bibinfo{year}{2019}).
\newblock \bibinfo{title}{What determines the x-ray intensity and duration of a
  solar flare?}
\newblock {\it \bibinfo{journal}{The Astrophysical Journal}\/},  {\it
  \bibinfo{volume}{874}\/}\bibinfo{issue}{(2)}, \bibinfo{pages}{157}.
\bibitem[{{Ryan} et~al.(2012){Ryan}, {Milligan}, {Gallagher}, {Dennis},
  {Tolbert}, {Schwartz} \& {Young}}]{Ryan2012}
\bibinfo{author}{{Ryan}, D.~F.}, \bibinfo{author}{{Milligan}, R.~O.},
  \bibinfo{author}{{Gallagher}, P.~T.}, \bibinfo{author}{{Dennis}, B.~R.},
  \bibinfo{author}{{Tolbert}, A.~K.}, \bibinfo{author}{{Schwartz}, R.~A.}, \&
  \bibinfo{author}{{Young}, C.~A.} (\bibinfo{year}{2012}).
\newblock \bibinfo{title}{{The Thermal Properties of Solar Flares over Three
  Solar Cycles Using GOES X-Ray Observations}}.
\newblock {\it \bibinfo{journal}{The Astrophysical Journal Supplement
  Series}\/},  {\it \bibinfo{volume}{202}\/}\bibinfo{issue}{(2)},
  \bibinfo{pages}{11}. \DOIprefix\doi{10.1088/0067-0049/202/2/11}.
  \href{http://arxiv.org/abs/1206.1005}{\tt arXiv:1206.1005}.
\bibitem[{{Sadykov} et~al.(2019){Sadykov}, {Kosovichev}, {Kitiashvili} \&
  {Frolov}}]{Sadykov2019}
\bibinfo{author}{{Sadykov}, V.~M.}, \bibinfo{author}{{Kosovichev}, A.~G.},
  \bibinfo{author}{{Kitiashvili}, I.~N.}, \& \bibinfo{author}{{Frolov}, A.}
  (\bibinfo{year}{2019}).
\newblock \bibinfo{title}{{Statistical Properties of Soft X-Ray Emission of
  Solar Flares}}.
\newblock {\it \bibinfo{journal}{The Astrophysical Journal}\/},  {\it
  \bibinfo{volume}{874}\/}\bibinfo{issue}{(1)}, \bibinfo{pages}{19}.
  \DOIprefix\doi{10.3847/1538-4357/ab06c3}.
  \href{http://arxiv.org/abs/1810.05610}{\tt arXiv:1810.05610}.
\bibitem[{Schwenn(2006)}]{schwenn2006space}
\bibinfo{author}{Schwenn, R.} (\bibinfo{year}{2006}).
\newblock \bibinfo{title}{Space weather: The solar perspective}.
\newblock {\it \bibinfo{journal}{Living reviews in solar physics}\/},  {\it
  \bibinfo{volume}{3}\/}\bibinfo{issue}{(1)}, \bibinfo{pages}{1--72}.
\bibitem[{{Smart} et~al.(2006){Smart}, {Shea} \& {McCracken}}]{Smart2006}
\bibinfo{author}{{Smart}, D.~F.}, \bibinfo{author}{{Shea}, M.~A.}, \&
  \bibinfo{author}{{McCracken}, K.~G.} (\bibinfo{year}{2006}).
\newblock \bibinfo{title}{{The Carrington event: Possible solar proton
  intensity time profile}}.
\newblock {\it \bibinfo{journal}{Advances in Space Research}\/},  {\it
  \bibinfo{volume}{38}\/}\bibinfo{issue}{(2)}, \bibinfo{pages}{215--225}.
  \DOIprefix\doi{10.1016/j.asr.2005.04.116}.
\bibitem[{Swalwell et~al.(2018)Swalwell, Dalla, Kahler, White, Ling, Viereck \&
  Veronig}]{swalwell2018}
\bibinfo{author}{Swalwell, B.}, \bibinfo{author}{Dalla, S.},
  \bibinfo{author}{Kahler, S.}, \bibinfo{author}{White, S.~M.},
  \bibinfo{author}{Ling, A.}, \bibinfo{author}{Viereck, R.}, \&
  \bibinfo{author}{Veronig, A.} (\bibinfo{year}{2018}).
\newblock \bibinfo{title}{The reported durations of goes soft x-ray flares in
  different solar cycles}.
\newblock {\it \bibinfo{journal}{Space Weather}\/},  {\it
  \bibinfo{volume}{16}\/}\bibinfo{issue}{(6)}, \bibinfo{pages}{660--666}.
\bibitem[{Temmer et~al.(2017)Temmer, Thalmann, Dissauer, Veronig, Tschernitz,
  Hinterreiter \& Rodriguez}]{temmer2017flare}
\bibinfo{author}{Temmer, M.}, \bibinfo{author}{Thalmann, J.~K.},
  \bibinfo{author}{Dissauer, K.}, \bibinfo{author}{Veronig, A.~M.},
  \bibinfo{author}{Tschernitz, J.}, \bibinfo{author}{Hinterreiter, J.}, \&
  \bibinfo{author}{Rodriguez, L.} (\bibinfo{year}{2017}).
\newblock \bibinfo{title}{On flare-cme characteristics from sun to earth
  combining remote-sensing image data with in situ measurements supported by
  modeling}.
\newblock {\it \bibinfo{journal}{Solar physics}\/},  {\it
  \bibinfo{volume}{292}\/}\bibinfo{issue}{(7)}, \bibinfo{pages}{1--22}.
\bibitem[{{Toriumi} \& {Wang}(2019)}]{Toriumi2019}
\bibinfo{author}{{Toriumi}, S.}, \& \bibinfo{author}{{Wang}, H.}
  (\bibinfo{year}{2019}).
\newblock \bibinfo{title}{{Flare-productive active regions}}.
\newblock {\it \bibinfo{journal}{Living Reviews in Solar Physics}\/},  {\it
  \bibinfo{volume}{16}\/}\bibinfo{issue}{(1)}, \bibinfo{pages}{3}.
  \DOIprefix\doi{10.1007/s41116-019-0019-7}.
  \href{http://arxiv.org/abs/1904.12027}{\tt arXiv:1904.12027}.
\bibitem[{T{\"o}r{\"o}k et~al.(2004)T{\"o}r{\"o}k, Kliem \& Titov}]{Tork2004}
\bibinfo{author}{T{\"o}r{\"o}k, T.}, \bibinfo{author}{Kliem, B.}, \&
  \bibinfo{author}{Titov, V.} (\bibinfo{year}{2004}).
\newblock \bibinfo{title}{Ideal kink instability of a magnetic loop
  equilibrium}.
\newblock {\it \bibinfo{journal}{Astronomy \& Astrophysics}\/},  {\it
  \bibinfo{volume}{413}\/}\bibinfo{issue}{(3)}, \bibinfo{pages}{L27--L30}.
\bibitem[{Verbeeck et~al.(2019)Verbeeck, Kraaikamp, Ryan \&
  Podladchikova}]{Verbeeck2019}
\bibinfo{author}{Verbeeck, C.}, \bibinfo{author}{Kraaikamp, E.},
  \bibinfo{author}{Ryan, D.~F.}, \& \bibinfo{author}{Podladchikova, O.}
  (\bibinfo{year}{2019}).
\newblock \bibinfo{title}{Solar flare distributions: Lognormal instead of power
  law?}
\newblock {\it \bibinfo{journal}{The Astrophysical Journal}\/},  {\it
  \bibinfo{volume}{884}\/}\bibinfo{issue}{(1)}, \bibinfo{pages}{50}.
\bibitem[{Veronig et~al.(2002)Veronig, Temmer, Hanslmeier, Otruba \&
  Messerotti}]{Veronig2002}
\bibinfo{author}{Veronig, A.}, \bibinfo{author}{Temmer, M.},
  \bibinfo{author}{Hanslmeier, A.}, \bibinfo{author}{Otruba, W.}, \&
  \bibinfo{author}{Messerotti, M.} (\bibinfo{year}{2002}).
\newblock \bibinfo{title}{Temporal aspects and frequency distributions of solar
  soft x-ray flares}.
\newblock {\it \bibinfo{journal}{Astronomy \& Astrophysics}\/},  {\it
  \bibinfo{volume}{382}\/}\bibinfo{issue}{(3)}, \bibinfo{pages}{1070--1080}.
\bibitem[{Viticchi{\'e} et~al.(2006)Viticchi{\'e}, Del~Moro \&
  Berrilli}]{Viticchie2006}
\bibinfo{author}{Viticchi{\'e}, B.}, \bibinfo{author}{Del~Moro, D.}, \&
  \bibinfo{author}{Berrilli, F.} (\bibinfo{year}{2006}).
\newblock \bibinfo{title}{Statistical properties of synthetic nanoflares}.
\newblock {\it \bibinfo{journal}{The Astrophysical Journal}\/},  {\it
  \bibinfo{volume}{652}\/}\bibinfo{issue}{(2)}, \bibinfo{pages}{1734--1739}.
\bibitem[{Vlahos et~al.(1995)Vlahos, Georgoulis, Kluiving \&
  Paschos}]{Vlahos1995}
\bibinfo{author}{Vlahos, L.}, \bibinfo{author}{Georgoulis, M.},
  \bibinfo{author}{Kluiving, R.}, \& \bibinfo{author}{Paschos, P.}
  (\bibinfo{year}{1995}).
\newblock \bibinfo{title}{The statistical flare.}
\newblock {\it \bibinfo{journal}{Astronomy and Astrophysics}\/},  {\it
  \bibinfo{volume}{299}\/}, \bibinfo{pages}{897--911}.
\bibitem[{{Wagner}(1988)}]{Wagner1988}
\bibinfo{author}{{Wagner}, W.~J.} (\bibinfo{year}{1988}).
\newblock \bibinfo{title}{{Observations of 1-8 {\r{A}} solar X-ray variability
  during solar cycle 21}}.
\newblock {\it \bibinfo{journal}{Advances in Space Research}\/},  {\it
  \bibinfo{volume}{8}\/}\bibinfo{issue}{(7)}, \bibinfo{pages}{67--76}.
  \DOIprefix\doi{10.1016/0273-1177(88)90173-1}.
\bibitem[{{Yashiro} et~al.(2006){Yashiro}, {Akiyama}, {Gopalswamy} \&
  {Howard}}]{Yashiro2006}
\bibinfo{author}{{Yashiro}, S.}, \bibinfo{author}{{Akiyama}, S.},
  \bibinfo{author}{{Gopalswamy}, N.}, \& \bibinfo{author}{{Howard}, R.~A.}
  (\bibinfo{year}{2006}).
\newblock \bibinfo{title}{{Different Power-Law Indices in the Frequency
  Distributions of Flares with and without Coronal Mass Ejections}}.
\newblock {\it \bibinfo{journal}{The Astrophysical Journal}\/},  {\it
  \bibinfo{volume}{650}\/}\bibinfo{issue}{(2)}, \bibinfo{pages}{L143--L146}.
  \DOIprefix\doi{10.1086/508876}.
  \href{http://arxiv.org/abs/astro-ph/0609197}{\tt arXiv:astro-ph/0609197}.
\bibitem[{Zhang \& Dere(2006)}]{zhang2006statistical}
\bibinfo{author}{Zhang, J.}, \& \bibinfo{author}{Dere, K.}
  (\bibinfo{year}{2006}).
\newblock \bibinfo{title}{A statistical study of main and residual
  accelerations of coronal mass ejections}.
\newblock {\it \bibinfo{journal}{The Astrophysical Journal}\/},  {\it
  \bibinfo{volume}{649}\/}\bibinfo{issue}{(2)}, \bibinfo{pages}{1100--1109}.
\bibitem[{Zhang et~al.(2001)Zhang, Dere, Howard, Kundu \&
  White}]{zhang2001temporal}
\bibinfo{author}{Zhang, J.}, \bibinfo{author}{Dere, K.},
  \bibinfo{author}{Howard, R.}, \bibinfo{author}{Kundu, M.}, \&
  \bibinfo{author}{White, S.} (\bibinfo{year}{2001}).
\newblock \bibinfo{title}{On the temporal relationship between coronal mass
  ejections and flares}.
\newblock {\it \bibinfo{journal}{The Astrophysical Journal}\/},  {\it
  \bibinfo{volume}{559}\/}\bibinfo{issue}{(1)}, \bibinfo{pages}{452--462}.

\end{thebibliography}

\clearpage
\appendix
\section{Comparison  between our algorithm and that of \cite{aschwanden_algo}}

\begin{table}[h]
    \small
    \centering
    \renewcommand*{\arraystretch}{1.4}

    \begin{tabular}{|m{20em}|m{20em}|}
    \hline
    \cite{aschwanden_algo} Algorithm &
    Proposed Algorithm \\
    \hline
    Data rebinning to obtain a 12s resolution & 
    Same technique \\
    \hline
    Elimination of data gaps and spike removal (Q parameter). The article does not specify how spikes are treated &
    We used the same criterium to find spikes. We tested data from year 1987, the year which presented the biggest number of spikes, and decided to remove portions of data affected by the problem, since they only represent a very small percentage of data \\
    \hline
    Definition of a minimum flare duration (5 bins or 60s) &
    We do not set a minimum flare duration, the maximum and minimum detection procedure used automatically solves this problem \\
    \hline
    Boxcar smoothing using a 21-points window. The article does not specify what kind of window has been used &
    We used a window of 21 bins and replaced the value at the center of the bin with the mean value over the window\\
    \hline
    All local maximum and minimum points obtained from the smoothed light curve are used as candidates for peak, start and end times &
    Even after the smoothing process, light curves still show small fluctuations. Not all maximum and minimum points are registered. A data point is considered maximum (minimum) only if its value is higher (lower) than the values of the two points on its left and of the two points on its right \\
    \hline
    The background is defined as the median flux in a time interval (60s) before the start of the event &
    Same definitions but we use the mean value instead of the median one \\
    \hline
    Definition of a noise parameter with a value $f_{noise} = 2\cdot10^{-8}$ W m\textsuperscript{-2} &
    Same definition \\
    \hline
    Threshold criteria to associate each starting point to an end point. The same ending point cannot be used for two different starting points. Intervals are therefore truncated at the start of another event &
    We allow the same ending point to be matched with different starting points and use the information to label multiple and single events. This is meaningful when considering the time duration of events and their energy \\
    \hline
    Given an interval the maximum value is registered as peak flux. All peak values are registered and sometimes the peak flux can be lower than the background, giving a negative absolute value $F = f_p -f_{BG} < 0$ &
    We apply a threshold criterium to the maximum point too. Since we require $f_p > f_{BG}+f_{treshold}$, with $f_{treshold}=f_{noise}$ we are excluding extremely small events which can be associated with data noise. At the same time the condition $F = f_p -f_{BG} > 0$ is guaranteed \\
    \hline
    Event duration is truncated at the beginning of the following one &
    With the information about overlapping events the original duration of the event can be found. Each event can be interpreted as a single event or, if overlapping in time with another, as part of a group \\
    \hline
    No information about the energy associated to the event &
    Our algorithm registers the energy associated to each event. Using the Multiple Id, single event energies can be added together to get the total energy \\
    \hline
    \end{tabular}

\end{table}

\end{document}